\documentstyle[12pt]{article}

\def\seCtion#1{\section{#1} \setcounter{equation}{0}}
\renewcommand\theequation{\ifnum\value{section}>0{\thesection.
\arabic{equation}}\fi}
\newcommand{\be}{\begin{equation}}
\newcommand{\ee}{\end{equation}}
\newcommand{\bea}{\begin{eqnarray}}
\newcommand{\eea}{\end{eqnarray}}

\newcommand{\bi}{\bibitem}
\newcommand{\la}{\label}

\begin{document}

\pagestyle{empty}
\begin{flushright}
\mbox{FIUN-CP-00/1}
\end{flushright}
\quad\\

\begin{center}
\Large {\bf Can large fermion chemical potentials suppress \\
the electroweak phase transition?}
\end{center}

\begin{center}
\vspace{0.8cm}
{\bf C. Quimbay$^{a,}$} \footnote{Associate researcher at Centro
Internacional de F\'{\i}sica, Santaf\'e de Bogot\'a, Colombia. \\
(a) carloqui@ciencias.ciencias.unal.edu.co \\
(b) johnmo@ciencias.ciencias.unal.edu.co \\
(c) rhurtado@ciencias.ciencias.unal.edu.co},
{\bf J. Morales}$^{b,~1}$ {\bf and R. Hurtado}$^{c,~1}$\\

{\it {Departamento de F\'{\i}sica}, Universidad Nacional de
Colombia\\ Ciudad Universitaria, Santaf\'e de Bogot\'a, D.C.,
Colombia }

\vspace{0.8cm}
November 8, 2000

\end{center}

\vspace{1.2cm}

\begin{abstract}
We calculate the critical temperature $(T_c$) of the electroweak
phase transition in the minimal standard model considering
simultaneously temperature ($T$) and fermion chemical potential
($\mu_f$) effects over the effective potential. The calculation is
performed in the one-loop approximation to the effective potential
at non-zero temperature using the real time formalism of the
thermal field theory. We show that it exists a fermion chemical
potential critical value ($\mu_f^c$) for which the Higgs boson
condensate vanishes at $T=0$. If $T$ and $\mu_f$ effects are
considered simultaneously, it is shown that for $\mu_f \geq \mu_f^c$
then $T_c^2 \leq 0$, implying that the electroweak phase transition
might not take place.

\vspace{0.8cm}


\end{abstract}

\newpage

\pagestyle{plain}


\seCtion{Introduction}

\hspace{3.0mm}
The original idea that spontaneously broken gauge symmetries might be
restored at high temperatures for elementary particles systems in
thermodynamic equilibrium was first presented by Kirzhnits and Linde
\cite{kirz}. They developed this idea for the case of field theories
with global gauge symmetries, while for theories with local gauge
symmetries it was done by Dolan and Jackiw \cite{dolan} and Weinberg
\cite{weinberg}. Later Linde found that the Higgs boson condensate,
responsible for the spontaneous symmetry breaking in gauge theories,
depends not only on temperature ($T$) but also on fermion chemical
potentials ($\mu_f$) \cite{Linde1, Linde2}. In reference
\cite{Linde1} Linde showed that in most gauge theories with neutral
currents an increase of $\mu_f$ leads to an increase of symmetry
breaking. This behaviour was shown explicitly for the abelian Higgs
model extended by the inclusion of fermions. For the Minimal Standard
Model (MSM) it was shown \cite{Linde1} that the electroweak symmetry
may not be restorated in the early Universe if an excess of neutrinos
over anti-neutrinos is sufficiently large at present. It should be
mentioned that Linde's results \cite{Linde1} were obtained in the
tree level approximation.

On the other hand, it has been established that the MSM is discarded
as the framework of a possible mechanism of baryogenesis that gives
a satisfactory explanation of the current Baryon Asymmetry of the
Universe (BAU) \cite{gavela}. This needling fact is due to the
smallness of CP violation effects when fermion damping rates are
included \cite{gavela}-\cite{wang}. Until now it has not been
presented a convincing mechanism of baryogenesis at temperatures
below the critical temperature $(T_c)$ of the electroweak phase
transition in the MSM \cite{kraus}. However it is plausible that the
BAU might have been generated at a $T$ above $T_c$ through a
mechanism of baryogenesis in which $B-L$ symmetry is conserved
\cite{dolgov}. By this reason we consider a scenario in which $B-L$
is conserved and the BAU has been generated at a $T$ above $T_c$
\cite{erdas}. In this scenario the number of particles can be
slightly larger than the number of anti-particles implying
non-vanishing fermion chemical potentials ($\mu_f \neq 0$).
Consequently the electroweak phase transition might take place
in a plasma characterized by an excess of particles over
anti-particles, i.e. $\mu_f \neq 0$.

The main goal of this work is to calculate the $T_c$ of the
electroweak phase transition in the MSM for a plasma characterized
by $\mu_f \neq 0$. This calculation is performed in the one-loop
approximation to the effective potential at finite temperature,
working in the the real time formalism of the thermal field theory
\cite{niemi}-\cite{land} and using the Feynman gauge. We first
calculate the $T$ and $\mu_f$ effects over the effective potential
and then we obtain the Higgs boson condensate dependence on $T$
and $\mu_f$. We show that at $T=0$ it exists a chemical potential
critical value $\mu_f^c$ for which the electroweak symmetry is
restored. If $T$ and $\mu_f$ effects are simultaneously considered
it is shown that for $\mu_f \geq \mu_f^c$ then $T_c^2 \leq 0$,
implying that the electroweak phase transition does not take place
in this scenario. In the limit $\mu_f = 0$ we reproduce the
expression for $T_c$ published in \cite{smilga}.

The one-loop approximation to the effective potential at finite
temperature has been calculated using  the method shown at
\cite{smilga}. Before performing our calculation for the MSM,
which will be done in section 3, we first consider the abelian Higgs
model extended by the inclusion of fermions in section 2. The study
of this toy model allows to investigate the dependence of the Higgs
boson condensate on $T$ and $\mu_f$, and to calculate the $T_c$ for
the case of an abelian gauge theory. Finally our conclusions are
presented in section 4.

\seCtion{The Abelian Higgs Model}

\hspace{3.0mm}
The abelian Higgs Model extended by the inclusion of a fermion field
$\psi$ and a Yukawa coupling term is given by the following
Lagrangian:
\bea
{\cal L}=-\frac{1}{4} \left( \partial_\mu A_\nu - \partial_\nu A_\mu
\right)^2+\left(\partial_\mu + ie A_\mu \right) \bar{\phi}
\left(\partial_\mu - ie A_\mu \right) \phi -\lambda \left(\bar{\phi}
\phi - \frac{b^2}{2 \lambda} \right)^2 \nonumber \\
+ \bar{\psi} i \gamma_\mu \left( \partial_\mu - ie A_\mu \right) \psi
+ Y_f \bar{\psi} \phi \psi + Y_f \bar \psi \bar \phi \psi, \la{1}
\eea
where the complex scalar field is $\phi=\frac{1}{\sqrt{2}}(H+i\eta)$
and $\bar{\phi}$ is its hermitic conjugate field, being  $H$ the
Higgs boson field, $\eta$ the Goldstone boson field and $Y_f$ the
Yukawa coupling constant between $\psi$ and $\phi$. We observe that
the potential at zero temperature $V(\phi, \bar{\phi})= \lambda
(\bar{\phi} \phi - \frac{b^2}{2 \lambda})^2$ has a minimum value
located at $<\phi \bar{\phi}>=H^2 + \eta^2 = \nu_o^2$, where
$\nu_o^2=\frac{b^2}{2\lambda}$.  The abelian gauge symmetry of the
model is spontaneously broken by the choice of a definite ground
state, for instance $H=\nu_o$ and $\eta=0$. This mechanism of
spontaneous symmetry breaking, usually called Higgs mechanism, acts
as the generator for the gauge boson and fermion masses of the
model described by ($\ref{1}$) in the framework of quantum field
theory at zero temperature.

For the statistical model in thermodynamic equilibrium, with the
same particle content and interactions as described by Lagrangian
($\ref {1}$), it is well known that for $\mu_f = 0$ the gauge
symmetry of the model at $T=0$ is spontaneously broken due to the
appearance of the Higgs boson condensate, i.e.
$<\phi \bar{\phi}>=\nu_o^2$. This statistical system presents a phase
transition at a critical temperature ($T_c$): for temperatures above
$T_c$ the system is in the symmetric phase, $\nu =0$, while for
temperatures below $T_c$ it is in the broken symmetry phase,
$\nu \neq 0$. However, as we will show later, this phase transition
might not occur if $T$ and $\mu_f$ effects are simultaneously
considered as $<\phi \bar{\phi}>$ vanishes for $\mu_f \geq \mu_f^c$
at $T \geq 0$.

In this section we consider a thermal medium constituted by fermions,
anti-fermions, Higgs bosons and gauge bosons, characterized by
$\mu_f \not = 0$, where $\mu_f$ is the chemical potential associated
to an excess of fermions over anti-fermions in the plasma. At finite
temperature and density the Feynman rules for vertices are the same
as those at $T=0$ and $\mu_f =0$, and the propagators in the Feynman
gauge for massless gauge boson $D_{\mu \nu} (p)$, massless scalars
$D(p)$ and massless fermions $S(p)$ are \cite{kobes}:

\bea
D_{\mu \nu}(p) &=& -g_{\mu \nu} \left[ \frac{1}{p^2+i\epsilon} - i
\Gamma_b (p) \right], \la{2} \\
D(p) &=& \frac{1}{p^2+i \epsilon} - i \Gamma_b (p), \la{3} \\
S(p) &=& \frac{\not p}{p^2+i \epsilon} + i \Gamma_f (p), \la{4}
\eea
where $p$ is the particle four-momentum. The plasma temperature $T$
is introduced through the $\Gamma_b (p)$ and $\Gamma_f (p)$ functions
given by:

\bea
\Gamma_b (p) &=& 2 \pi \delta (p^2) n_b (p), \la{5} \\
\Gamma_f (p) &=& 2 \pi \delta (p^2) n_f (p), \la{6}
\eea
with
\bea
n_b (p) &=& \frac{1}{e^{(p \cdot u)/T}-1}, \la{7} \\
n_f (p) &=& \theta(p \cdot u)n_f^- (p) + \theta(-p \cdot u)n_f^+ (p)
\la{8}
\eea
where $n_b (p)$ is the Bose-Einstein distribution function. The
Fermi-Dirac distribution functions for fermions $n_f^- (p)$ and for
anti-fermions $n_f^+ (p)$ are:

\bea
n_f^{\pm} (p)=\frac{1}{e^{(p \cdot u \pm \mu_f)/T}+1}. \la{9}
\eea
In the distribution functions ($\ref{7}$) and ($\ref{8}$) $u^\alpha$
is the four-velocity of the center-mass frame of the dense plasma
with $u^\alpha u_\alpha =1$.

If fermion density effects are not considered the change of
$<\phi \bar{\phi}>$, from $\nu_o^2$ at $T=0$ to zero at $T=T_c$, is
due to the temperature corrections to the effective potential of the
complex scalar field. Now we calculate this effective potential
considering simultaneously $T$ and $\mu_f$ effects following the same
procedure as shown in \cite{smilga}. We first calculate the
polarization operator $\Pi_{\beta}^{\phi}$ of the complex scalar
field $\phi$ at zero external momentum:

\bea
\Pi_{\beta}^{\phi}(0) = \left [ \frac{\partial^2 V_{\beta}(\phi,
\bar{\phi})} {\partial \phi \partial \bar{\phi}} \right ]_{\phi =0}.
\la{10}
\eea
The one-loop diagrams which contribute to $\Pi_{\beta}^{\phi}$ are
shown in Fig. 1. The contributes of these diagrams to
$\Pi_{\beta}^{\phi}$ at leading order in $T^2$ and $\mu_f^2$ are:

\bea
\Pi_{(a)}(0) &=& \frac{1}{3}\lambda T^2,  \\
\Pi_{(b)}(0) &=& \frac{1}{3}e^2 T^2,  \\
\Pi_{(c)}(0) &=& -\frac{1}{12}e^2 T^2,  \\
\Pi_{(d)}(0) &=& \frac{Y_f^2}{2}\left[\frac{T^2}{3}+\frac{\mu_f^2}
{\pi^2} \right]. \la{11}
\eea
We observe that the $\mu_f$ effects over the effective potential at
finite temperature is due to the contribute of diagram $(d)$ shown
in Fig. 1., which includes a fermion propagator. The full effective
potential is:

\bea
V(\phi, \bar{\phi})+V_{\beta}(\phi, \bar{\phi}) = \lambda (\phi
\bar{\phi}- \nu_o^2)^2 + \left[ \frac{1}{12}(4\lambda+3e^2+
2Y_f^2)T^2+ Y_f^2 \frac{\mu_f^2}{2\pi^2} \right] \phi \bar{\phi}.
\la{15}
\eea
The Higgs boson condensate dependence on $T$ and $\mu_f$ is given by:

\bea
\nu^2(T,\mu_f)=\nu_o^2-\frac{1}{6} \left(1+\frac{3e^2}{4\lambda}+
\frac{Y_f^2}{2\lambda} \right)T^2 -\frac{1}{4\pi^2}\frac{Y_f^2}
{\lambda} \mu_f^2. \la{16}
\eea
For the purpose of reproducing known results we put $T=0$ in
($\ref{16}$) and we obtain:

\bea
\nu^2(0,\mu_f)=\nu_o^2-\frac{1}{4\pi^2}\frac{Y_f^2}{\lambda}\mu_f^2.
\la{17}
\eea
As it is possible to observe in ($\ref{17}$) $\nu(0,\mu_f)$ turns to
be zero at a fermion chemical potential critical value $\mu_f^c$
given by:

\bea 
\mu_f^c=\sqrt{2}\pi \frac{b}{Y_f}. \la{18}
\eea
This result is in accordance with equation (5.3) of reference
\cite{Linde2} implying that, if temperature effects are not included,
a second order phase transition with symmetry restoration takes place
\cite{Linde2}. The Higgs boson condensate ($\ref{16}$) turns to be
zero at:

\bea
T=T_c=2 \nu_o \sqrt{\frac{6\lambda  - 3(Y_f^2 \mu_f^2/2 \pi^2
\nu_o^2)} {4\lambda + 3e^2 + 2Y_f^2}}. \la{19}
\eea
We observe that if fermions are not introduced in the Higgs model
Lagrangian ($\ref{1}$), i.e. $Y_f=0$ in ($\ref{16}$), then
($\ref{19}$) reduces to:

\bea
T_c=2\nu_o \sqrt{6 \lambda / (4\lambda + 3e^2)}, \la{20}
\eea
in agreement with \cite{smilga}. Expression ($\ref{19}$) is very
interesting because it shows the effect of $\mu_f$ on the $T_c$
value. We note from ($\ref{19}$) that for:

\bea
\mu_f = \mu_f^c = \pi \nu_o \frac{m_H}{m_f}, \la{21}
\eea
where $m_H^2=2 \lambda \nu_o^2$ and $m_f^2=Y_f^2 \nu_o^2/2$, then
$T_c = 0$. This means that if we consider simultaneously $T$
and $\mu_f$ effects, and if $\mu_f \geq \mu_f^c$, then
$T_c^2 \leq 0$. This implies that in the above scenario a phase
transition can not take place at any temperature. In other words
if $\mu_f$ is sufficiently large the gauge symmetry of the Higgs
model extended by the inclusion of fermions is not spontaneously
broken for $T  \geq 0$.


\seCtion{The Minimal Standard Model}

\hspace{3.0mm}
In this section we will calculate the $T_c$ of the electroweak phase 
transition in the MSM. The scenario considered corresponds to an
electroweak plasma in thermodynamical equilibrium characterized by
unknown non-vanishing fermion chemical potentials. We consider for
quarks $\mu_u \not = \mu_d \not = \mu_c \not = ... \not = 0$ and for
charged leptons $\mu_e \not = \mu_{\mu} \not = \mu_{\tau} \not = 0$.
Following a similar procedure as the one of section 2 we obtain the
effective potential at non-zero temperature:

\bea
V_{\beta}(\phi, \bar{\phi}) = \left[ 2 \lambda + \frac{3 g^2}{4}
+ \frac{g'^2}{4} + \sum_{q}Y_q^2 + \frac{1}{3} \sum_{l}Y_l^2 \right]
\frac{T^2}{4} \phi \bar{\phi} \nonumber \\
+ \left[ \frac{3}{4\pi^2} \sum_{q} Y_q^2 \mu_q^2 + \frac{1}{4\pi^2}
\sum_{l} Y_l^2 \mu_l^2 \right] \phi \bar{\phi}, \la{22}
\eea
where the only novelty is the contribute from the different fermion
species through the Feynman diagrams as the one of figure (1d). With
this result the Higgs boson condensate depends on $T$ and $\mu_f$ as:

\bea
\nu(T, \mu_f)^2 = \nu_o^2 - \left[1 + \frac{3 g^2}{8\lambda}
+ \frac{g'^2}{8\lambda} + \frac{1}{2\lambda}\sum_{q}Y_q^2 +
\frac{1}{6\lambda} \sum_{l}Y_l^2 \right] \frac{T^2}{4}\nonumber \\
- \frac{1}{8\pi^2 \lambda} \left[ 3 \sum_{q} Y_q^2 \mu_q^2 + \sum_{l}
Y_l^2 \mu_l^2 \right]. \la{23}
\eea
The electroweak phase transition occurs at 

\bea
T=T_c(\mu_f)= 2 \nu_o \frac{ \left[ 1  -  \frac{3}{2 \pi^2} \sum_{q}
\frac{m_q^2}{m_H^2} \frac{\mu_q^2}{\nu_o^2} - \frac{1}{2\pi^2}
\sum_{l} \frac{m_l^2}{m_H^2} \frac{\mu_l^2}{\nu_o^2} \right] ^{1/2}}
{ \left[1 + \frac{3 g^2}{8\lambda} + \frac{g'^2}{8\lambda} +
2 \sum_{q} \frac{m_q^2}{m_H^2} + \frac{2}{3} \sum_{l}
\frac{m_l^2}{m_H^2} \right] ^{1/2} }, \la{24}
\eea
where we have used the notation $m_f^2= Y_q^2\nu_o^2/2$ and $m_H^2 =
2 \lambda \nu_o^2$. We observe that if we do not consider fermion
contributes to the effective potential, i.e. $m_f = 0$ in
($\ref{24}$), we obtain:

\bea
T_c= 2\nu_o \left[1 + \frac{3 g^2}{8\lambda} +
\frac{g'^2}{8\lambda} \right]^{-1/2},  \la{25}
\eea
in agreement with \cite{kapusta}. If we put $\mu_f = 0$ in
($\ref{24}$), we obtain:

\bea
T_c(0)= 2\nu_o \left[1 + \frac{3 g^2}{8\lambda} + \frac{g'^2}
{8\lambda} + 2 \sum_{q} \frac{m_q^2}{m_H^2} + \frac{2}{3}
\sum_{l} \frac{m_l^2}{m_H^2} \right]^{-1/2}, \la{26}
\eea
in accordance with \cite{smilga}.

It is clear from ($\ref{24}$) that if:

\bea
\frac{3}{2 \pi^2} \sum_{q} \frac{m_q^2}{m_H^2} \frac{\mu_q^2}
{\nu_o^2} + \frac{1}{2\pi^2} \sum_{l} \frac{m_l^2}{m_H^2}
\frac{\mu_l^2}{\nu_o^2} = 1, \la{27}
\eea
then $T_c = 0$. The contributes to ($\ref{24}$) coming from the
different fermion species are proportional to $m_f^2/m_H^2$. The
experimental lower bound on the Higgs boson mass, $M_H =$107 GeV
\cite{bock}, allows to conclude that the main contribute in
($\ref{24}$) is due to the top quark mass ($m_t$), and it is
possible to write in good approximation ($\ref{27}$) as:

\bea
\frac{3}{2 \pi^2} \frac{m_t^2}{m_H^2} \frac{\mu_t^2}{\nu_o^2} \sim 1.
\la{28}
\eea

Taking $m_t=174$ GeV, $\nu_o=246$ GeV and $m_H=110$ GeV in
($\ref{28}$), we obtain a chemical potential critical value:

\be
\mu_t^{c} \sim 400~~{\rm GeV}. \la{29}
\ee

From ($\ref{24}$) it can be stated that if $\mu_t \geq \mu_t^{c}$
then $T_c^2 \leq 0$, therefore the electroweak phase transition can
not take place in the above scenario.

For the case $\mu_t < \mu_t^c$ we observe from ($\ref{24}$) that
$T_c(\mu_t) < T_c(0)$, being $T_c(0) \sim 100$ GeV the critical
temperature value of the electroweak phase transition for the case
$\mu_t = 0$. The $T_c(0)$ value can be obtained from ($\ref{26}$)
taking the same inputs as in ($\ref{29}$). We note that the $\mu_f$
values associated to the different fermion species of the MSM in the
expression ($\ref{24}$) are unknown parameters.


\section{Conclusions}

\hspace{3.0mm}
We have consider a $B-L$ conserving thermal medium in thermodynamic
equilibrium in which the Baryon Asymmetry of the Universe (BAU) has
been generated at a temperature $T$ above the critical temperature
$T_c$ of the electroweak phase transition in the Minimal Standard
Model (MSM). In this scenario the number of particles can be
slightly larger than the number of anti-particles implying
non-vanishing fermion chemical potentials $\mu_f \neq 0$. We have
calculated the $T_c$ in the MSM for a thermal medium characterized
by $\mu_f \neq 0$. The calculation was performed in the one-loop
approximation to the effective potential at finite temperature,
working in the real time formalism of the thermal field theory and
using the Feynman gauge. To calculate the effective potential we
have followed the procedure shown at \cite{smilga}.

We have calculated the $T$ and $\mu_f$ effects over the effective
potential, and the Higgs boson condensate dependence on $T$ and
$\mu_f$. We have shown that at $T=0$ it exists a chemical potential
critical value $\mu_f^c$ for which the electroweak symmetry is
restored. If $T$ and $\mu_f$ effects are simultaneously considered
it was shown that for $\mu_f \geq \mu_f^c$ then $T_c^2 \leq 0$,
implying that the electroweak phase transition does not take place
in this scenario. In the limit $\mu_f = 0$ we have reproduced the
$T_c$ expression published in \cite{smilga}.

On the other hand we have shown for $\mu_f < \mu_f^c$ that
$T_c(\mu_f) < T_c(0)$, being $T_c(0)$ the critical temperature of
the electroweak phase transition for the case $\mu_f = 0$. The main
non-vanishing fermion chemical potential effect over the electroweak
phase transition is to lower the $T_c$ value respect to the case
$\mu_f = 0$. We have assumed that $\mu_{f_i} \not = 0$ for all the
$f_i$ quark and charged lepton flavours. The values of the different
$\mu_{f_i}$ are unknown and we argue that their values could be
obtained through a mechanism of baryogenesis generating the current
BAU at a $T$ above $T_c$.


\section*{Acknowledgements}
This work was supported by COLCIENCIAS (Colombia) and by Universidad
Nacional de Colombia under research grant DIB 803629.



\newpage

\begin{figure}[1]
\let\picnaturalsize=N
\def\picsize{3in}
\def\picfilename{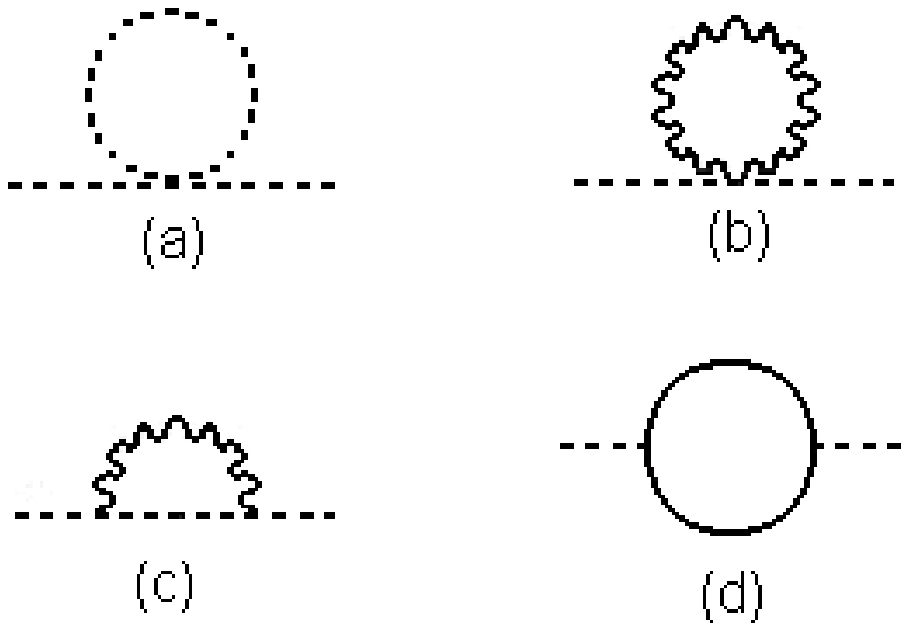}
\caption{One-loop diagrams contributing to the effective potential
at finite temperature.}
\label{Fig. 1.}
\end{figure}

\end{document}